\documentclass{iopart}
\usepackage{epsfig}
\newcommand{\beq}{\begin{equation}}
\newcommand{\eeq}{\end{equation}}
\newcommand{\bea}{\begin{eqnarray}}
\newcommand{\eea}{\end{eqnarray}}
\newcommand{\bce}{\begin{center}}
\newcommand{\ece}{\end{center}}
\newcommand{\eg}{{\it e.g.}}
\newcommand{\ie}{{\it i.e.}}
\def\lsim{\mathrel{\rlap{\lower4pt\hbox{\hskip1pt$\sim$}}
    \raise1pt\hbox{$<$}}}         %less than or approx. symbol
\def\gsim{\mathrel{\rlap{\lower4pt\hbox{\hskip1pt$\sim$}}
    \raise1pt\hbox{$>$}}}         %greater than or approx. symbol

\begin{document}

\title{Theoretical Overview on (Hidden) Charm in High-Energy Heavy-Ion 
       Collisions} 

\author{Ralf Rapp\dag\footnote[3]{email: rapp@nordita.dk} and 
Loic Grandchamp\ddag\footnote[4]{email: loic@tonic.physics.sunysb.edu}
}

\address{\dag\ NORDITA, Blegdamsvej 17, DK-2100 Copenhagen, Denmark} 

\address{\ddag\ Department of Physics, SUNY Stony Brook, NY 11794-3800,
          U.S.A. and IPN Lyon, 69622 Villeurbanne Cedex, France}

\begin{abstract}
Recent developments in the theoretical evaluation of charmonium 
production in ultrarelativistic heavy-ion collisions (URHIC's) are 
discussed. 
In particular, the consequences of equilibrium properties of open  
and hidden charm states -- accessible, {\it e.g.}, in QCD lattice 
gauge calculations -- are assessed. These include abundances as well 
as  formation and dissociation rates of charmonia in both hadronic 
and quark-gluon matter.   
\end{abstract}

%Uncomment for PACS numbers title message
%\pacs{00.00, 20.00, 42.10}

% Uncomment for Submitted to journal title message
%\submitto{\JPG}

% Comment out if separate title page not required
%\maketitle

%%%%%%%%%%%%%%%%%%%%%%%%%%%%%%%%%%%%%%%%%%%%%%%%%%%%%%%%%%%%%%%%%%%%%%%%
\section{Introduction}
\label{sec_intro}
%%%%%%%%%%%%%%%%%%%%%%%%%%%%%%%%%%%%%%%%%%%%%%%%%%%%%%%%%%%%%%%%%%%%%%%%
Hadrons containing charm quarks play a special role in 
ultrarelativistic heavy-ion physics~\cite{Shu78,MS86,KP87,BO87,Svet88,
RBS88,GGJ88,HLZ88,Khar98,GH99,Vogt99}. 
On the one hand, the charm ($c$) quark carries a mass, 
$m_c$$\simeq$1.3~GeV/c$^2$, large enough for its production to be 
dominated by primordial nucleon-nucleon ($N$-$N$) 
collisions~\cite{LMW95}; on the other hand, $m_c$ is small enough to 
allow for significant interactions with surrounding light quarks and 
gluons, thus providing a rather direct probe of the hot and/or dense 
matter created in central collisions of heavy nuclei ($A$-$A$). 
This particularly applies for $c$-$\bar c$ bound states (charmonia): 
based on the production systematics in $p$-$p$ reactions, where they 
amount to  $\sim$2\% of the total charm ($c\bar c$) yield, 
charmonia have been predicted to be suppressed beyond standard nuclear 
effects if a deconfined medium is formed in $A$-$A$ 
collisions~\cite{MS86}.  
At the same time, reinteractions of $c$-quarks 
imply approach towards thermalization, and, in principle, the presence 
of the backward channel in charmonium dissociation reactions,  
\beq
X_1 + J/\psi  \rightleftharpoons  X_2 + c + \bar c \ (D + \bar D) \ . 
\label{react}
\eeq   
To identify the relevant processes in a heavy-ion reaction, several 
questions arise, \eg,
\begin{itemize}  
\item 
What are the inelastic charmonium cross sections, and corresponding
equilibra\-tion times,  in the Quark-Gluon Plasma (QGP) and hadronic 
phase?
\item 
What are the in-medium properties (mass and width) of the charmonium 
and open charm states in both phases?
\item
How does the primordially produced number of charmonium states 
compare to thermal equilibrium abundances?
\item
What is the significance of the critical temperature $T_c$ in the 
previous items?
\item
Are there large corrections to the expected open charm content of
the system, and do $c$-quarks ($D$-mesons) thermalize? 
\end{itemize}

In this talk we will address (some of) these issues as follows.  
In Sec.~\ref{sec_open} we briefly recall the basic 
features of open charm production, including new insights from
recent RHIC data. In Sec.~\ref{sec_hidden}, we first assess  
in-medium properties of open and hidden charm states from recent 
lattice QCD results. Second, we elaborate on their possible 
consequences on the description of charmonium evolution in URIHC's 
in both QGP and hadronic phases. The main tool will be rate equations 
with their main ingredients -- initial conditions, relaxation times 
and equilibrium limits -- being discussed in some detail. Third,    
we will comment on statistical hadronization models, in particular 
how in-medium modifications of charmed hadrons may affect their results. 
In Sec.~\ref{sec_model}, we confront various approaches to available
data, and in Sec.~\ref{sec_concl} we conclude.

%%%%%%%%%%%%%%%%%%%%%%%%%%%%%%%%%%%%%%%%%%%%%%%%%%%%%%%%%%%%%%%%%%%%%%%%
\section{Open Charm Production}
\label{sec_open}
%%%%%%%%%%%%%%%%%%%%%%%%%%%%%%%%%%%%%%%%%%%%%%%%%%%%%%%%%%%%%%%%%%%%%%%%
The production of $c\bar c$ pairs in heavy-ion reactions is expected to 
behave as a hard process, \ie, scale with the number of primordial
$N$-$N$ collisions. While $p$-$A$ data follow this 
expectation~\cite{e789,na50imr}, only indirect measurements via 
semileptonic decays, $c (\bar c) \to l^+ (l^-) X$  (within charmed 
hadrons), are available for the $A$-$A$ case so far. 

In $Pb$-$Pb$ at SPS energy, $\sqrt{s}$=17.3~AGeV, 
NA50~\cite{na50imr} found an enhancement of intermediate-mass 
({\it IM}, $M_{\mu\mu}$=1.5-2.5~GeV) dimuon pairs over the expected 
Drell-Yan and open charm sources, gradually 
increasing with centrality up to a factor of $\sim$2.5 in most central 
events.  This excess is well reproduced if an increase in open charm 
production by a factor of $\sim$3.5 is postulated. However, underlying 
mechanisms for such an increase are not easily conceivable. In fact, 
it has been found~\cite{RS00,GKP00,KGS02} that {\em thermal} dimuon 
radiation from a fireball with reasonable initial temperatures, 
$T_i$=200-250~MeV, can, maybe more naturally, account for the {\it IM} 
dimuon excess. The experimental resolution of this problem is expected 
from the NA60 experiment~\cite{na60}.     

In $Au$-$Au$ at RHIC energy, $\sqrt{s}$=130,200~AGeV, 
single-electron ($e^\pm$) transverse-momen\-tum ($p_t$) spectra have 
been measured at various centralities~\cite{phenix130e,phenix200e}. 
After subtraction of light-hadron decay 
sources, the remaining spectra can be accounted for by "standard" 
charm-production extrapolated from $N$-$N$ event generators.
On the one hand, this limits the possibility for an  
appreciable charm enhancement. On the other hand, it is somewhat 
surprising that even for central collisions the spectral shape is
reasonably well reproduced without any reinteractions, as one would 
naively expect a softening if charm quarks (partially) thermalize. 
However, thermalization also implies that $c$-quarks participate in the 
collective matter expansion. In Ref.~\cite{Bat02} it has been shown
that the current PHENIX data~\cite{phenix130e,phenix200e}
for "charm-like" $e^\pm$ are also consistent 
with the assumption of complete $c$-quark thermalization {\it and} 
collective flow. As discussed below, $c$-quark reinteractions have 
important consequences for charmonium production; the obvious
observable to disentangle the two extremes is the $c$-quark
{\it elliptic} flow~\cite{Bat02}.   
       
%%%%%%%%%%%%%%%%%%%%%%%%%%%%%%%%%%%%%%%%%%%%%%%%%%%%%%%%%%%%%%%%%%%%%%%%
\section{Charmonium in QCD Matter}
\label{sec_hidden}
%%%%%%%%%%%%%%%%%%%%%%%%%%%%%%%%%%%%%%%%%%%%%%%%%%%%%%%%%%%%%%%%%%%%%%%%

%%%%%%%%%%%%%%%%%%%%%%%%%%%%%%%%%%%%%%%%%%%%%%%%%%%%%%%%%%%%%%%%%%%%%%%%
\subsection{Lattice QCD Results}
\label{ssec_lat}
\begin{figure}[!t]
\begin{center}
\epsfig{file=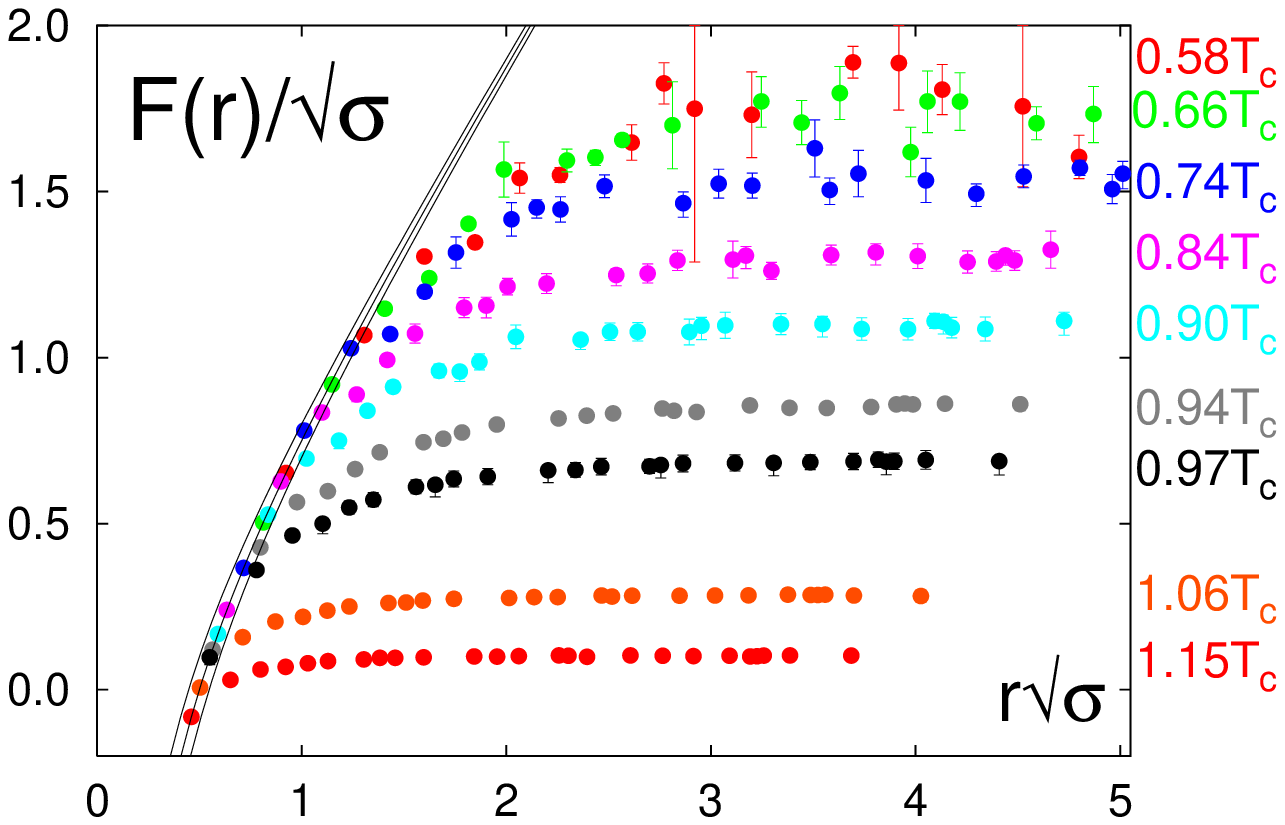,width=5.65cm}
\hspace{0.5cm}
\epsfig{file=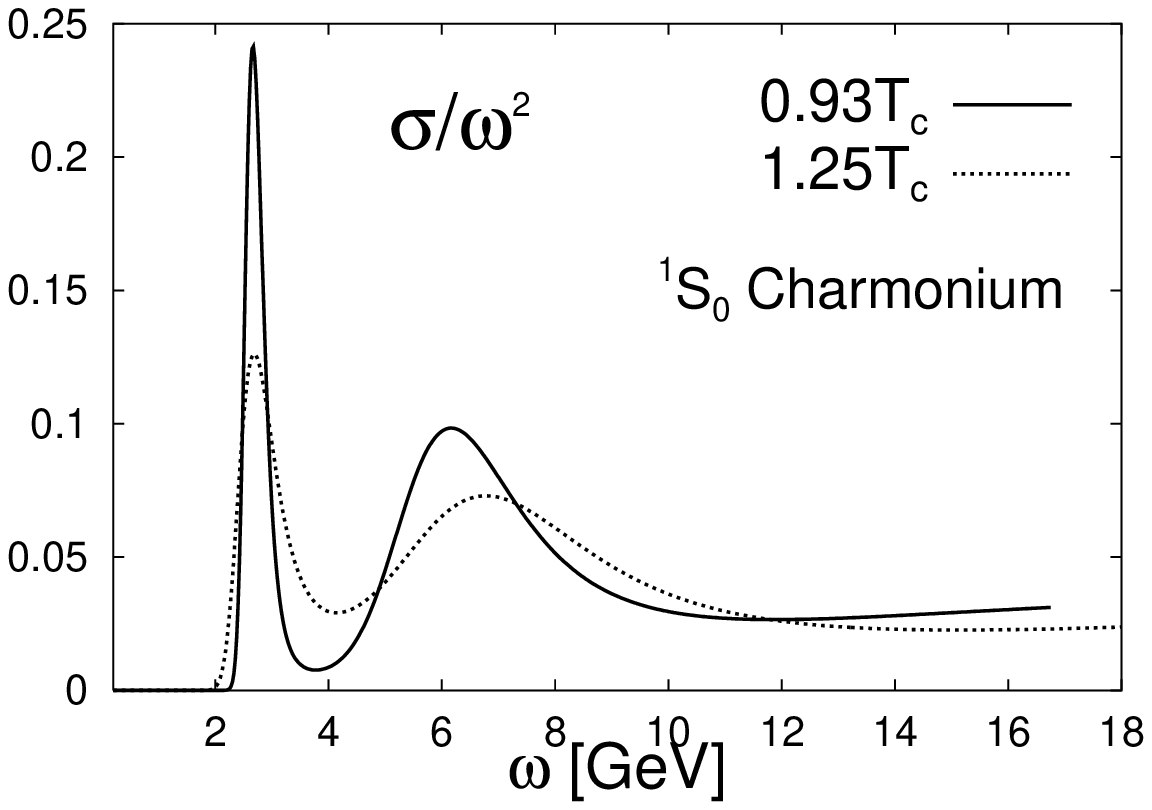,width=6.7cm}
\end{center}
\caption{Left panel: free energy of a heavy-quark pair from unquenched
($N_f$=2+1) lattice calculations~\protect\cite{KLP01} as a function of
$Q$-$\bar Q$ separation $r$ for various temperatures $T$. Right panel: 
quenched lattice results~\protect\cite{Datta02} for the
$\eta_c$ spectral function above and below $T_c$.}
\label{fig_lat}
\end{figure}

%%%%%%%%%%%%%%%%%%%%%%%%%%%%%%%%%%%%%%%%%%%%%%%%%%%%%%%%%%%%%%%%%%%%%%%%
Lattice QCD calculations for the finite-$T$ free energy of a 
heavy-quark pair~\cite{KLP01}, 
\beq
F_{Q\bar Q}(r,T)=V_{Q\bar Q}(r,T)-T S_{Q\bar Q}(r,T) \ , 
\eeq
including dynamical fermions, are displayed in the left panel of 
Fig.~\ref{fig_lat}. 
The decrease of the asymptotic plateau value,
$F_{Q\bar Q}(T,r\to\infty)$,  
can be attributed to an in-medium reduction of the open charm
threshold in the heavy-quark potential, $V_{Q\bar Q}$, and has been 
used to infer "dissociation" temperatures
$T_{diss}$$<$$T_c$ for $\psi'$ and $\chi$ states, as well as 
$T_{diss}$$\simeq$$1.1~T_c$ for the $J/\psi$~\cite{DPS01}.   
It should be emphasized that the temperature dependence of the plateau
is remarkably continuous, even across $T_c$; this has interesting 
implications which we will return to below.   
If bound states move above the threshold for spontaneous decay, it 
does, however, not necessarily mean that they disappear from the 
spectrum, but they can rather survive as resonances\footnote{note 
that, in a heat bath, the distinction between bound state and resonance
becomes immaterial; all information (mass and width) is encoded in the 
spectral function.}. This is indeed found in two recent (quenched) 
lattice calculations~\cite{Datta02,UNM02}, where reconstructed $\eta_c$ and 
$J/\psi$ spectral functions exhibit well-defined states up to 
$T$$\simeq$1.5~$T_c$, see right panel of Fig.~\ref{fig_lat}.  
The masses appear to be essentially $T$-independent (quite
contrary to the open charm states), whereas the widths, 
$\Gamma_{\eta_c,\psi}$, increase significantly, to about 0.2-0.5~GeV 
slightly above $T_c$~\cite{Datta02,UNM02}. This is not much different 
from lifetimes obtained using gluon-induced $J/\psi$ dissociation
cross sections~\cite{GR01} at comparable temperatures.

%%%%%%%%%%%%%%%%%%%%%%%%%%%%%%%%%%%%%%%%%%%%%%%%%%%%%%%%%%%%%%%%%%%%%%%%
\subsection{Rate Equations in the Quark-Gluon Plasma}
\label{ssec_rate}
\begin{figure}[!t]
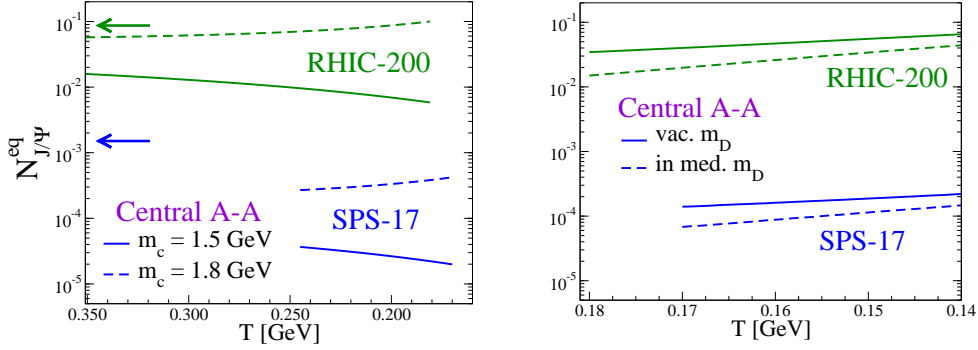

\begin{center}
\epsfig{file=charm_qgp_eq.eps,width=6.2cm}
\hspace{0.75cm}
\epsfig{file=charm_hg_eq.eps,width=5.7cm}
\end{center}
\caption{Left panel: Equilibrium $J/\psi$ abundances (assuming 
$m_\psi(T)\equiv m_\psi^{vac}$) in an isotropic,
adiabatically expanding Quark-Gluon Plasma at fixed (anti-) charm-quark 
number, $N_{c\bar c}$, resembling conditions at RHIC and SPS for two 
values of the charm-quark mass. The arrows mark the numbers, 
$N_\psi^{prim}$, of $J/\psi$'s produced in primordial (hard) $N$-$N$
collisions (as extrapolated from elementary $p$-$p$ reactions).  \\ 
Right panel: $J/\psi$ equilibrium numbers in the hadronic phase
(under equivalent conditions as in the left panel) 
using free and in-medium charmed hadron masses.}  
\label{fig_Neq}
\end{figure}

%%%%%%%%%%%%%%%%%%%%%%%%%%%%%%%%%%%%%%%%%%%%%%%%%%%%%%%%%%%%%%%%%%%%%%%%
Let us now turn to the description of the time evolution of $J/\psi$
abundances relevant for heavy-ion reactions, first focusing on the  
QGP phase. A corresponding rate equation for the number of 
$J/\psi$ mesons in a system (fireball) of volume 
$V_{FB}$, $N_\psi(\tau)=n_\psi(T) V_{FB}(\tau)$ ($n_\psi(T)$: 
$J/\psi$ density), is of the form
\beq
\frac{d}{d\tau} N_\psi = - N_\psi L(\tau) + G(\tau) 
= -\frac{1}{\tau_\psi} \left[ N_\psi(\tau) -N_\psi^{eq}(\tau) \right]
 \ ,  
\label{rate}
\eeq 
which applies as long as a well-defined 
$J/\psi$ state exists, \ie, 
$\tau_\psi^{-1}(T)$=$\Gamma_\psi(T)$$\ll$$m_\psi$. The second equality
in Eq.~(\ref{rate}) follows under the simplifying assumption that the 
surrounding partons (including the charm quarks) are in thermal 
equilibrium. The charm-quark and $J/\psi$ equilibrium densities are 
then given by 
\bea
n_c(T,\gamma_c)&=&n_{\bar c}(T,\gamma_c)= 6\gamma_c 
\int \frac{d^3q}{(2\pi)^3} f^c(m_c;T)
\label{nc}
\\
n^{eq}_\psi(T,\gamma_c)&=&  3\gamma_c^2 
\int \frac{d^3q}{(2\pi)^3} f^\psi(m_\psi;T) 
 \ . 
\label{npsi}
\eea
Here we have already incorporated the possibility of chemical 
{\em off}-equilibrium by introducing the charm-quark fugacity $\gamma_c$. 
In practice, the latter will be matched to the total number, 
$N_{c\bar c}$, of $c\bar c$ pairs produced in hard (primordial) $N$-$N$ 
collisions~\cite{pbm00},  
\beq
N_{c\bar c}= \frac{1}{2} N_{op} \ \frac{I_1(N_{op})}{I_0(N_{op})}  \
+ \ V_{FB}(T) \ \gamma_c(T)^2 \sum\limits_{X=\eta_c,\psi,\dots}  n_X(T)
\label{Ncc}
\eeq
with $N_{op}$=$2~V_{FB}(T)~\gamma_c(T)~n_c(T)$.
%% (the second term -- hidden charm, $N_{hid}$ -- can usually be neglected).
Fig.~\ref{fig_Neq} (left panel) shows equilibrium $J/\psi$ numbers in 
the QGP under conditions expected for central $Pb$-$Pb$/$Au$-$Au$ 
collisions at SPS/RHIC. One observes a large sensitivity to the (in-medium) 
charm-quark mass, with larger values for $m_c$ requiring a larger 
fugacity $\gamma_c$ (at fixed $N_{c\bar c}$), and thus entailing a 
{\em larger} equilibrium abundance of charmonium states. 

\begin{figure}[!t]
\begin{center}
\epsfig{file=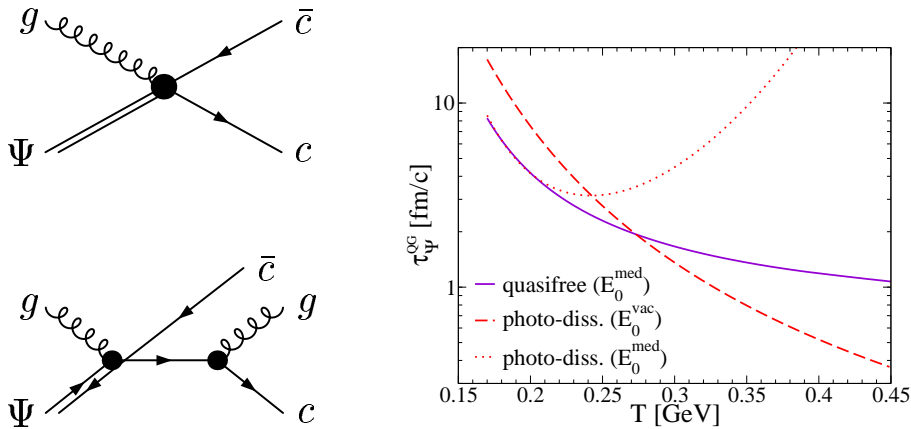,width=5.0cm}
\hspace{0.5cm}
\epsfig{file=tau_qgp.eps,width=6.7cm}
\end{center}
\vspace{-0.5cm}
\caption{Left panel: parton-induced $J/\psi$ dissociation processes
(upper panel: the equivalent of photodissociation induced by gluons; 
lower panel: quasifree destruction);  
Right panel: Corresponding $J/\psi$ relaxation times.  }
\label{fig_tauqgp}
\end{figure}

Besides the equilibrium $J/\psi$ number the other essential quantity
figuring into Eq.~(\ref{rate}) is the (chemical) relaxation time 
$\tau_\psi$ (related to processes given in Eq.~(\ref{react})).  
In a perturbative framework, it is determined in terms 
of parton-induced $J/\psi$ breakup (inelastic) cross sections, 
$\sigma_{\psi}^{inel}$, via convolution over thermal parton distributions 
$f^{i}$,  
\beq
\tau_\psi^{-1} = \int  \frac{d^3k}{(2\pi)^3} \ \sum\limits_{i=q,\bar q,g}
 f^{i}(k;T) \ \sigma_{i\psi}^{inel} \ . 
\eeq 
The use of the standard gluodissociation cross 
section~\cite{Shu78,BP79}, depicted in the upper left panel of 
Fig.~\ref{fig_tauqgp}, leads to relaxation times 
$\tau_\psi$$\ge$10~fm/c for temperatures $T$$\le$200~MeV, steeply 
falling to below 1~fm/c at $T$$\ge$300~MeV. 
If, however, according to the lattice data in Fig.~\ref{fig_lat}, 
the threshold energy of the outgoing open charm states is substantially 
reduced and approaches the $J/\psi$ mass, this process becomes 
kinematically unfavo\-rable, see dotted line in right panel of
Fig.~\ref{fig_tauqgp}. 
Under these circumstances, "quasifree" 
dissociation~\cite{GR01}, which is characterized by an extra outgoing 
parton (lower left panel in Fig.~\ref{fig_tauqgp}), is much more 
efficient, cf.~solid line in the right panel of Fig.~\ref{fig_tauqgp}.     

Nonperturbative approaches to the dissociation of $J/\psi$'s in a QGP 
often involve the notion that also "$D$"-meson states survive the 
phase transition~\cite{HLZ88,Bla02}; \eg, in constituent-quark based 
pictures~\cite{Bla02}, $D$-mesons exhibit the "Mott" effect, \ie,  
become broad resonances in the $c$-$\bar q$ continuum above $T_c$, 
thus opening phase space for $J/\psi\to "D" + "{\bar D}"$ decays. 
      
The final ingredient needed for an actual solution of the rate
equation is the initial condition, $N_\psi^0$, usually taken from 
hard production systematics (modified by initial-state nuclear 
absorption to account for "pre-equilibrium" dynamics).
One can then specify under which conditions the gain term in 
Eq.~(\ref{rate}) can be neglected in applications to heavy-ion reactions, 
namely: either (i) $\tau_\psi\gg \tau_{QGP}$ ($\tau_{QGP}$: lifetime of 
the QGP phase) and  $N_\psi^0 \gg N_\psi^{eq}$ or (ii) no $J/\psi$ 
state is supported in the QGP. 
The arrow markers in the left panel of Fig.~\ref{fig_Neq} indicate  
how the expected primordial production numbers of $J/\psi$'s compare 
to the thermal equilibrium levels. Whereas at SPS energies
$N_\psi^{eq} \ll N_\psi^{prim}$, this no longer holds for RHIC 
energies, \ie, a neglection of the gain term appears to be
justified at SPS, but this is not obvious at RHIC.

%%%%%%%%%%%%%%%%%%%%%%%%%%%%%%%%%%%%%%%%%%%%%%%%%%%%%%%%%%%%%%%%%%%%%%%%
\subsection{Hadronic Matter: Statistical Production and in-Medium 
Effects}
\label{ssec_hg}
%%%%%%%%%%%%%%%%%%%%%%%%%%%%%%%%%%%%%%%%%%%%%%%%%%%%%%%%%%%%%%%%%%%%%%%%
Based on the success of thermal models to describe the production 
of hadrons containing $u$, $d$ and $s$ quarks~\cite{BRS03}, it
has been suggested~\cite{pbm00} that also charmed hadrons  
form by statistical hadronization (of pre-existing $c$, $\bar c$ quarks) 
according to thermal weights at $T_c$.  
This, in particular, has been applied to charmonium 
states~\cite{pbm00,Goren01,ABRS03} by employing 
Eqs.~(\ref{npsi}) and (\ref{Ncc}) but with the fugacity 
$\gamma_c$ determined by the hadronic open charm states,  
$N_{op} = V_{FB}(T_c)~\gamma_c~\sum_\alpha n_\alpha(T,\mu_B)$, where 
$\alpha$= $D$, $\bar D$, $D^*$, $\bar D^*$, $\Lambda_c$, \dots, 
runs over all known open charm hadrons, and $V_{FB}(T_c)$ is the 
hadronic fireball volume after completion of the phase transition
(see also Ref.~\cite{GG99}).     

In the language of the rate Eq.~(\ref{rate}), the statistical 
hadronization approach cor\-responds to having reached complete 
thermal and (relative) chemical equilibrium at $T_c$.  Interactions in 
the subsequent hadronic phase may still alter the charmonium 
abundances, as determined by inelastic hadronic reactions of type
$\pi J/\psi \leftrightarrow D \bar D^* + {\rm c.c.}$, 
$\rho J/\psi \leftrightarrow D \bar D$,  etc.,  which have been 
intensely investigated, mostly with focus on the dissociation 
direction. 
\begin{figure}
\begin{center}
\epsfig{file=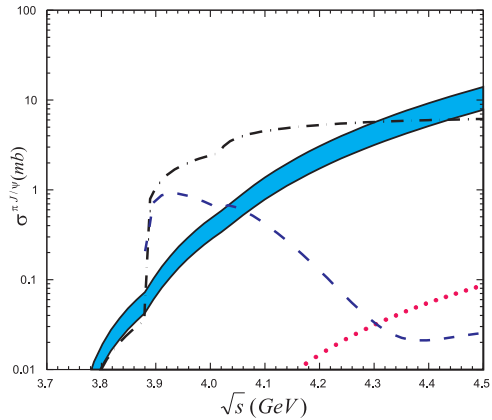,width=6.55cm}
\end{center}
\caption{$\pi~J/\psi \to D\bar D, D^*\bar D, D\bar D^*$ dissociation 
cross sections in various approaches; 
band: QCD sum rules~\protect\cite{Dura03}, 
dotted line: short-distance QCD~\protect\cite{KS94}, 
dashed-dotted line: meson-exchange~\protect\cite{OSL01}, dashed line: 
nonrelativistic constituent quark model~\protect\cite{Wong01}.} 
\label{fig_had}
\end{figure}
An example of the current status is shown in Fig.~\ref{fig_had}, taken
from Ref.~\cite{Dura03}: the band, which is the main result of that 
work, represents the pion-induced dissociation cross-section as assessed
within a QCD sum rule analysis. It is in reasonable agree\-ment with 
evaluations from both constituent-quark~\cite{MBQ95,Wong01} and  
meson-exchange models~\cite{MM98,HG01,LK00,OSL01,NNR01} 
(once soft form factors with cutoffs $\Lambda$$\le$1~GeV are included), 
but significantly above earlier short-distance 
QCD calculations~\cite{KS94} based on gluodissociation~\cite{BP79}
folded with the (very soft) gluon distribution functions within light 
hadrons. $\rho$-~\cite{Wong01,HG01} and nucleon-induced~\cite{LKL01} 
break-up cross sections tend to be somewhat larger. But even under 
rather extreme assumptions, \eg, an average cross section of 1~mb and 
a total hadron density $n_{had}$=3$n_0$$\simeq$0.5~fm$^{-3}$, the 
chemical relaxation time, 
$\tau_\psi$=$[\langle\sigma v\rangle n_{had}]^{-1}$$\simeq$30-40~fm/c,  
is well below typical hadronic fireball lifetimes. 

The situation may change if additional medium effects are present, most
notably reduced $D$-meson masses, as indicated by the lattice results 
in the left panel of Fig.~\ref{fig_lat}, QCD sum rule 
estimates~\cite{Hay00} or model calculations~\cite{Tsu99}.  
In a constituent quark picture, such modifications are attributed to 
the light quarks losing their mass, thus directly reflecting on 
(partial) chiral symmetry restoration. From a practical point of view, 
the lowering of the $D\bar D$ threshold has several consequences, \eg:
(i) excited charmonia may move above threshold (well) below 
$T_c$~\cite{DPS01},
(ii) increased phase space accelerates the inelastic reactions 
(although wave function nodes in excited charmonia can lead to 
strong deviations from naive extrapolations~\cite{FLS02}) 
(iii) the equilibrium abundance of charmonia in hadronic matter 
decreases (at fixed $N_{c\bar c}$). The latter effect 
is illustrated in the right panel of Fig.~\ref{fig_Neq}, and is 
completely analogous to varying $m_c$ in the QGP, see discussion 
after Eq.~(\ref{Ncc}). Even a moderate reduction of the $D$-meson 
masses by 80~MeV at $T_c$=170~MeV lowers the $J/\psi$ equilibrium 
level by a factor of 2.  
The other remarkable feature of the hadronic equilibrium $J/\psi$ 
numbers is that they increase with decreasing temperature -- again 
a consequence of the increasing $\gamma_c$ at fixed $N_{c\bar c}$  
together with $M_\psi < 2m_D$\footnote{In the academic limit $T\to 0$, 
all $c\bar c$ pairs would be in the lowest possible state, \ie, 
$\eta_c$ mesons!}.
%, \ie it is energetically 
%favorable to store a $c\bar c$ pair in a $J\psi$ rather than in a $D\bar D$ 
%pair (note that for the QGP curves in the left panel of 
%Fig.~\ref{fig_Neq}, $M_\psi < 2m_c$ for $m_c$=1.8GeV, but $M_\psi > 2m_c$ 
%for $m_c$=1.5GeV, implyiong a decreasing curve). 
For heavy-ion reactions this implies that, should the $J/\psi$ number
reach equilibrium at or around $T_c$, subsequent hadronic
reactions would favor additional formation over dissociation (this 
could be even more relevant for excited charmonium states which only 
exist in hadronic matter). Such a behavior has indeed been observed in
kinetic~\cite{BMR00} and transport~\cite{Zhang02,BCS03} model calculations.   

%%%%%%%%%%%%%%%%%%%%%%%%%%%%%%%%%%%%%%%%%%%%%%%%%%%%%%%%%%%%%%%%%%%%%%%%
\subsection{Continuity?}
\label{ssec_cont}
%%%%%%%%%%%%%%%%%%%%%%%%%%%%%%%%%%%%%%%%%%%%%%%%%%%%%%%%%%%%%%%%%%%%%%%%
Before we turn to quantitative model comparisons to heavy-ion data, 
let us reiterate ramifications of the lattice QCD results~\cite{KLP01} 
for the heavy-quark free energy (left panel of Fig.~\ref{fig_lat})  
on the $J/\psi$ evolution within the rate-equation framework.   
As mentioned above, the asymptotic plateau, $F_{c\bar c}(T,r\to\infty)$, 
suggests a gradual decrease of the open charm threshold with 
increasing temperature~\cite{DPS01}, which is surprisingly continuous
even when passing through the phase transition.
At the same time, the $J/\psi$ mass is essentially constant. The latter
implies that, in thermal and chemical equilibrium 
($\gamma_c$=1), the $J/\psi$ {\em density}, Eq.~(\ref{npsi}), 
only depends on temperature and thus carries no notion of phase 
transition dynamics. Of course, in a finite system, the latent heat 
released in going from quark-gluon to hadronic degrees of freedom would 
imply the total $J/\psi$ number to increase 
according to the volume expansion at $T_c$.  
However, this is not the situation envisaged in a heavy-ion 
reaction, where, in addition, one expects the total charm-quark number, 
$N_{c\bar c}$, to be approximately constant, encoded in the chemical 
off-equilibrium parameter $\gamma_c$$\ne$1. Hence, through  
Eqs.~(\ref{npsi}) and (\ref{Ncc}), $J/\psi$ abundances are coupled to  
$N_{c\bar c}$ {\em and} spectral distributions of open-charm states 
(figuring into the densities), parametrically as 
$N_\psi\propto n_\psi/(V_{FB} n_{op}^2)$ with $n_{op}$=$2n_c(m_c^*,T)$ 
or $\sum_{\alpha=D,D^*,...} n_\alpha(m_\alpha^*,T)$,  
corresponding to QGP or hadron gas (HG), respectively.   
It furthermore seems conceivable that the open-charm spectrum -- not 
only the lowest level -- changes rather continuously through $T_c$, and
therefore $n_{op}^{QGP}$$\simeq$$n_{op}^{HG}$ (\eg, with in-medium 
charm-quark masses $m_c^*$$>$$m_c$=1.15-1.35GeV/c$^2$ due to 
heavy-quark thermal correlation energies in the QGP).   
In this case, the equilibrium charmonium numbers {\em decrease} during 
the phase transition according to the volume increase at $T_c$, quite
contrary to the chemical-equilibrium scenario ($\gamma_c$$\equiv$1).

%%%%%%%%%%%%%%%%%%%%%%%%%%%%%%%%%%%%%%%%%%%%%%%%%%%%%%%%%%%%%%%%%%%%%%%%
\section{Model Approaches at SPS and RHIC}
\label{sec_model}
\begin{figure}
\begin{center}
\epsfig{file=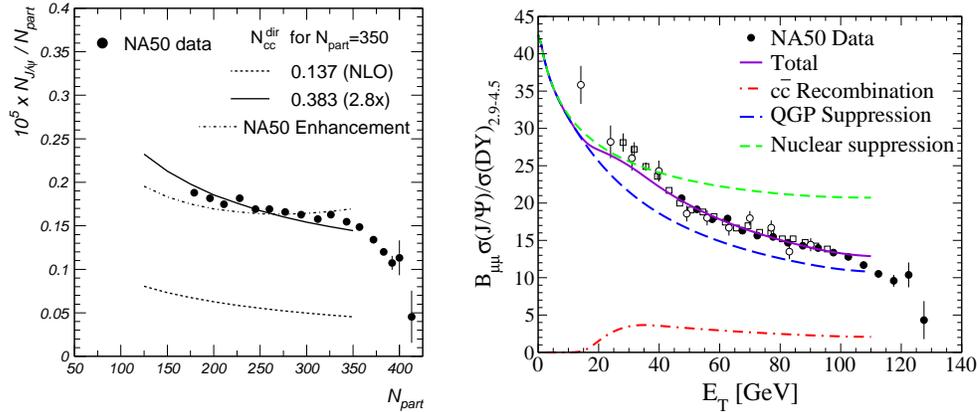,width=5.9cm}
\hspace{0.2cm}
\epsfig{file=jpsi-sps.eps,width=6.7cm}
\end{center}
\caption{Recent model approaches to NA50 $J/\psi$ 
data~\protect\cite{na50-00} from $Pb$-$Pb$ collisions at SPS 
($\sqrt{s_{NN}}$=17.3~GeV). 
Left panel: statistical coalescence~\protect\cite{ABRS03};
right panel: two-component model~\protect\cite{GR01}.}
\label{fig_sps}
\end{figure}

%%%%%%%%%%%%%%%%%%%%%%%%%%%%%%%%%%%%%%%%%%%%%%%%%%%%%%%%%%%%%%%%%%%%%%%%
At SPS energies, the interpretation of the NA50 
data~\cite{na50-00,na50-03} is still under debate. On the one hand, 
"comover" models~\cite{Cap00} with a constant dissociation cross 
section of $\sigma_{co}^{inel}$$\simeq$1~mb are able to explain the 
observed suppression.  One should note, however, that the underlying 
comover densities reached in the early stages of a collision --
albeit not necessarily thermal -- 
correspond to energy densities well above critical ones 
obtained from lattice QCD. On the other hand, the recent 
idea~\cite{pbm00,Goren01} of statistical coalescence of $c$ and 
$\bar c$ quarks at $T_c$ has stimulated further activity. One of its 
strongholds is the $\psi'/\psi$ ratio~\cite{pbm00}, which 
experimentally~\cite{na50-98} tends to saturate at the thermal value 
in suf\-ficiently central $A$-$A$ collisions. To describe 
the $J/\psi$ data by statistical production alone (assuming all 
primordial $J/\psi$'s to be suppressed or not to form), requires an 
open charm enhancement by a factor of $\sim$3 (in line with the excess 
inferred from $IM$ dimuon spectra, cf.~the discussion in 
Sec.~\ref{sec_open}), see left panel of Fig.~\ref{fig_sps}. 
In Ref.~\cite{GR01} a two-com\-ponent approach, combining suppression 
mechanisms (nuclear, QGP and hadronic) with statistical production at 
$T_c$, has been suggested which allows to incorporate peripheral $A$-$A$ 
and $p$-$A$ collisions and refrains from introducing an anomalous charm 
enhancement. In this case, QGP suppression is identified as the key 
feature  at SPS, see right panel of Fig.~\ref{fig_sps}.    

\begin{figure}
\begin{center}
\epsfig{file=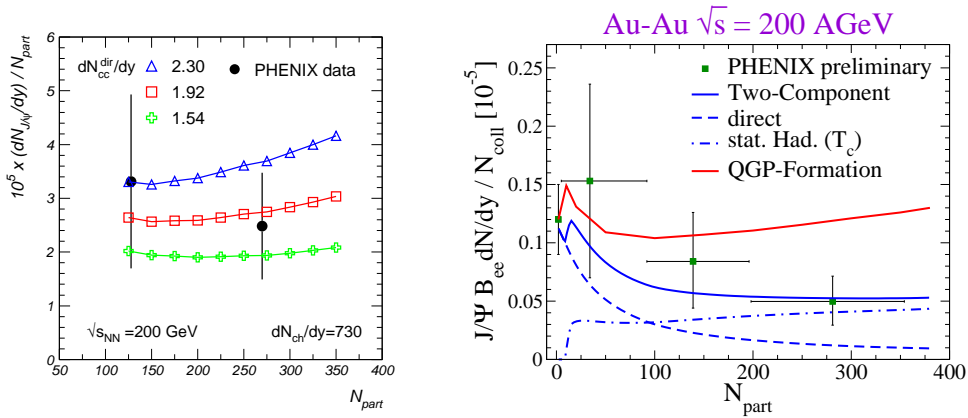,width=6.2cm}
\epsfig{file=jpsi-models-rhic.eps,width=6.7cm}
\end{center}
\caption{Model predictions for $J/\psi$ production at RHIC 
($\sqrt{s_{NN}}$=200~GeV) compared to preliminary PHENIX 
data~\protect\cite{phenix-psi}. Right panel: statistical
hadroni\-zation~\protect\cite{ABRS03}; left panel: dynamical approaches
(lower solid line: two-component model~\protect\cite{GR01}, upper solid 
line: yields from solving rate equations in a 
QGP~\protect\cite{TSR01}).}
\label{fig_rhic}
\end{figure}

Let us finally confront predictions for RHIC energies with 
(preliminary) data from PHENIX~\cite{phenix-psi}. The left panel of 
Fig.~\ref{fig_rhic} shows the statistical model 
yields~\cite{pbm00,ABRS03}, illustrating inherent uncertainties due to 
the thus far not accurately known open-charm cross section. In the 
right panel of Fig.~\ref{fig_rhic}, the results of the two-component 
model~\cite{GR01,GR03} indicate that for (semi-) central collisions at
RHIC the thermal production at $T_c$ indeed dominates over the 
primordial contribution, as the latter is subject to strong QGP 
suppression~\cite{Vogt99}. This has interesting consequences for the 
excitation function from SPS to RHIC energies~\cite{GR01}. The upper 
solid line in the right panel of Fig.~\ref{fig_rhic} is obtained from 
calculations~\cite{TSR01} solving the full rate Eq.~(\ref{rate}), 
using the gluodissociation cross section with a vacuum $J/\psi$ mass 
and width. Although $\Gamma_\psi$ is probably increased, the idea of a 
$J/\psi$ state in the QGP {\em is} supported by lattice QCD, recall
the right panel of Fig.~\ref{fig_lat}. The $J/\psi$ formation
in this calculation (which does not invoke a hadronic phase) 
indeed occurs preferentially in the early QGP stages which accounts 
for most of the difference to the 2-component model (where, in turn,
$J/\psi$ formation in the QGP has been neglected). Predictions 
in line with the current data were also made within a multiphase 
transport model in Ref.~\cite{Zhang02}, see also Ref.~\cite{BCS03} 
for a recent analysis. In both these calculations,  
secondary $J/\psi$ formation, \ie, the backward reaction in 
Eq.~(\ref{react}), again plays a decisive role.  
In Ref.~\cite{Cap03} it has been argued, that a reduced nuclear 
absorption could be consistent with the data without invoking any 
regeneration processes. 

%%%%%%%%%%%%%%%%%%%%%%%%%%%%%%%%%%%%%%%%%%%%%%%%%%%%%%%%%%%%%%%%%%%%%%%%
\section{Conclusion}
\label{sec_concl}
%%%%%%%%%%%%%%%%%%%%%%%%%%%%%%%%%%%%%%%%%%%%%%%%%%%%%%%%%%%%%%%%%%%%%%%%
About 20 years after the first ideas on utilizing charmonium states 
as a probe of hot and dense QCD matter in heavy-ion collisions, 
important insights continue to emerge. 
In addition to experimental data,  
lattice QCD calculations are now providing valuable information
for theoretical model approaches, most notably the survival of (some) 
charmonia as resonance states in the QGP as encoded in pertinent
spectral functions, and the smooth 
behavior of open charm thresholds through both hadronic and early
QGP phases, across $T_c$. We attempted to infer
consequences thereof in the heavy-ion environment, \eg,
the necessity to include charmonium formation reactions,   
the behavior of $J/\psi$ equilibrium levels across the 
hadronization transition and its interplay with in-medium
effects on open charm states.  
The current high-energy frontier at 
RHIC (and future LHC) implies, for the first time, multiple production
of $c\bar c$ pairs, as well as more favorable conditions for their 
reinteractions. If charm quarks indeed thermalize in the course of
an $A$-$A$ collision, the equilibrium level of charmonium states being 
comparable to primordial production, necessarily entails 
significant regeneration.  Whether such a regime has already been 
reached with RHIC energies
cannot be decided from the present $J/\psi$ and single-$e^\pm$ data. 
Future measurements in the upcoming RHIC runs (including open charm 
observables and possibly excited charmonia), in connection with further 
theoretical developments, will hopefully hold (some) answers.  

\vspace{0.5cm}

\noindent{\bf Acknowledgement} \\
One of us (RR) thanks the organizers of the conference for the 
invitation to this very informative and pleasant meeting. 
We are grateful to D. Blaschke, C.M. Ko and 
E.V. Shuryak for stimulating discussions. This work was supported 
in part by the U.S. DOE under contract no. DE-FG02-88ER40388.

\end{document}